\documentclass[twocolumn,showpacs,preprintnumbers,amsmath,amssymb]{revtex4}

\usepackage{slashed}

\def\tg{\tilde{g}}

\newcommand{\be}{\begin{equation}}
\newcommand{\ee}{\end{equation}}
\newcommand{\bea}{\begin{eqnarray}}
\newcommand{\eea}{\end{eqnarray}}
\newcommand{\ba}{\begin{array}}
\newcommand{\ea}{\end{array}}

\def\bbox{{\,\lower0.9pt\vbox{\hrule \hbox{\vrule height 0.2 cm
\hskip 0.2 cm \vrule height 0.2 cm}\hrule}\,}}
\newcommand{\dsl}{\pa \kern-0.5em /}

\usepackage{graphicx}
\usepackage{dcolumn}
\usepackage{bm}
\makeatletter\let\corresponds\@undefined\makeatother
\usepackage{mathabx}

\begin{document}

\preprint{LMU-ASC 08/12, MPP-2012-7}

\title{A geometric action for non-geometric fluxes}

\author{David Andriot$^{a}$, Olaf Hohm$^{a}$, Magdalena Larfors$^{a}$, Dieter L\"ust$^{a,b}$, Peter Patalong$^{a,b}$} 
\affiliation{%
\vskip0.15cm
$^{a}$Arnold-Sommerfeld-Center for Theoretical Physics, Fakult\"at f\"ur Physik, Ludwig-Maximilians-Universit\"at M\"unchen, \\Theresienstra\ss e 37, 80333 M\"unchen, Germany. \\ 
\\ \vskip-0.6cm 
$^{b}$Max-Planck-Institut f\"ur Physik, \\F\"ohringer Ring 6, 80805 M\"unchen, Germany. 
}%

\begin{abstract}
We give a geometrical interpretation of the non-geometric $Q$ and $R$ fluxes. To this end we consider  
double field theory 
in a formulation that is related to the conventional one by a field redefinition taking the form of a T-duality inversion. 
The $R$ flux is a tensor under diffeomorphisms and satisfies a non-trivial Bianchi identity. 
The $Q$ flux can be viewed as part of a connection that 
covariantizes the winding derivatives with respect to diffeomorphisms. We give a higher-dimensional action with a kinetic term 
for the $R$ flux and a `dual' Einstein--Hilbert term containing the connection $Q$.

\end{abstract}

\maketitle

String theory is a consistent theory in ten dimensions, and it is 
important to understand which four-dimensional theories it can give rise to.
Phenomenologically, the so-called gauged supergravities are particularly 
interesting; the gaugings fix compactification moduli and can allow 
de Sitter vacua. Some gauged supergravities can be 
obtained through flux compactifications of string theory or its low-energy supergravity limit.      
For instance, in the Neveu--Schwarz--Neveu--Schwarz (NS-NS) low-energy effective action for 
superstring theory, 
 \bea\label{original}
  S  =  \int dx \sqrt{-g}e^{-2\phi}\left[{\cal R}(g)+4(\partial\phi)^2-\frac{1}{12}H^{ijk}H_{ijk}\right]\,,  
 \eea
where $H_{ijk}=3\partial_{[i}b_{jk]}$ is the field strength of the  
NS-NS two-form $b_{ij}$, we may give a vacuum expectation value 
to this three-form. This results in massive or gauged supergravities  
in four dimensions.  It is known, however, that there are 
more gauged supergravities in four dimensions that cannot be 
obtained through any conventional (flux) compactification. 
This is part of the motivation to consider \textit{non-geometric}
fluxes that are related to the conventional fluxes 
via T-duality \cite{Shelton:2005cf}. The T-duality rules 
suggest a chain 
 \bea\label{chain}
  H_{abc}\rightarrow f^{a}{}_{bc}\rightarrow Q_{c}{}^{ab}\rightarrow R^{abc}\;.
 \eea
Here, $H$ is the conventional three-form field strength discussed above, 
and $f^{a}{}_{bc}=-2e_{[b}{}^{m}e_{c]}{}^{n}\partial_{m} e_{n}{}^{a}$, 
with the vielbein $e_{m}{}^{a}$, are the so-called `geometric fluxes',  
which are related to the Levi-Civita spin connection.  The geometrical 
interpretation of the non-geometric fluxes $Q$ and $R$ 
and the formulation of a higher-dimensional action 
will be  the content of this letter. A more detailed exposition will 
appear elsewhere \cite{ahllp12}. 

Recently, some of us have performed a field redefinition 
in ten-dimensional supergravity and obtained an action that 
captures part of the $Q$ flux \cite{allp11}.  
Introducing the field ${\cal E}_{ij}=g_{ij}+b_{ij}$ 
encoding the spacetime metric and the NS-NS 2-form, 
this field redefinition takes the form of a T-duality inversion in all ten 
directions:
 \bea\label{tildeE}
    {\cal E} \rightarrow  \tilde{\cal E}  \equiv  {\cal E}^{-1}\;, 
    \quad \tilde{\cal E}^{ij} =  \tilde{g}^{ij}+\beta^{ij}\;,     
 \eea 
where $\beta\rightarrow -\beta$ compared to \cite{allp11}.   
Being the inverse of ${\cal E}$, $\tilde{\cal E}$ naturally carries upper indices, 
satisfying $\tilde{\cal E}^{ik}{\cal E}_{kj}=\delta^{i}{}_{j}$, and  
we have decomposed $\tilde{\cal E}^{ij}$ into 
its symmetric and antisymmetric part.  
This gives rise to (the inverse of) a new metric $\tilde{g}_{ij}$
and an antisymmetric bivector $\beta^{ij}$. Moreover, a new dilaton was introduced via
 \bea\label{dilintro}
  e^{-2\phi} \sqrt{-g}=e^{-2\tilde{\phi}} \sqrt{-\tg} \ ,  
 \eea
where $\tg=\det{\tilde{g}_{ij}}$.    
In \cite{allp11} the simplifying assumption 
has been made that $\beta^{ij}\partial_{j}=0$ when acting on 
arbitrary fields. The action (\ref{original}) then reads in the new 
variables, up to total derivatives, 
 \bea\label{Qaction}
  S  =  \int dx \sqrt{-\tilde{g}}e^{-2\tilde{\phi}}\left[{\cal R}(\tilde{g})+4(\partial\tilde{\phi})^2-\frac{1}{2}|Q|^2\right]\,,  
 \eea
where 
 \bea\label{Q}
  Q_{m}{}^{nk} = \partial_{m}\beta^{nk}\;.
 \eea
Being a partial derivative of a bivector, $Q$ is not a tensor, 
but it can be checked that its failure to transform covariantly becomes 
irrelevant upon using the simplifying assumption  $\beta^{ij}\partial_{j}=0$.
Let us now relax this assumption and consider the full field redefinition  
of (\ref{original}). We then find that this gives rise to a new term involving 
part of the $R$ flux, which is a tensor, but the role of $Q$ in the full action is somewhat obscure. 
In particular, there does not appear to be a covariant tensor that reduces to 
(\ref{Q}) upon using the assumption. We will show in this letter that the 
proper geometric interpretation of $Q$ becomes apparent once we consider the 
field variables $\tilde{g}_{ij}$ and $\beta^{ij}$ in the context of double field theory (DFT), 
where $Q$ will play the role of a connection rather than a tensor. 
This allows us to write a geometric action for $Q$ and $R$ fluxes. 

We begin by reviewing DFT, which is an approach to make T-duality a manifest symmetry 
by doubling the coordinates at the level of the 
spacetime action for string theory \cite{Hull:2009mi,Hohm:2010jy,Hohm:2010pp}.
(See also earlier work by Siegel and Tseytlin \cite{Siegel:1993th,Tseytlin:1990va}.)
In addition to the usual coordinates $x^{i}$ associated to momentum 
modes there are new coordinates $\tilde{x}_i$ associated to winding modes, 
which combine into a fundamental vector $X^{M}=(\tilde{x}_i,x^{i})$ under 
the T-duality group $O(10,10)$. Although the coordinates are formally doubled 
we have to impose  the `strong constraint' 
 \bea
  \eta^{MN}\partial_{M}\partial_{N}  =  0 \;, \qquad 
   \eta^{MN} = \begin{pmatrix} 0 &   {\bf 1} \\ {\bf 1} & 0 \end{pmatrix}\;,
 \eea
where $\eta^{MN}$ denotes the $O(10,10)$ invariant metric and 
$\partial_{M}=(\tilde{\partial}^i,\partial_i)$. This constraint holds   
on arbitrary fields, parameters and their products, so that 
in particular
 \bea\label{strng}
  \partial_i A\,\tilde{\partial}^i B+  \tilde{\partial}^i  A \, \partial_i B = 0 \;, 
 \eea
for any $A,B$.  
This constraint implies that for any solution the fields depend only on half of the 
coordinates.   The double field theory formulation that is most convenient 
for our present purposes is based on the field ${\cal E}_{ij}$ discussed above 
and a dilaton density $d$, which is related to the scalar dilaton $\phi$ 
via the field redefinition $e^{-2d}=e^{-2\phi}\sqrt{-g}$ \cite{Hohm:2010jy}. 
Its action reduces  to (\ref{original}) 
upon setting $\tilde{\partial}^i=0$. 

DFT is invariant under a `generalized diffeomorphism' symmetry parametrized by an
$O(10,10)$ vector parameter $\xi^{M}=(\tilde{\xi}_i,\xi^i)$. For $\tilde{\partial}^i=0$ 
this reduces to conventional general coordinate transformations
$x^{i}\rightarrow x^{i}-\xi^{i}(x)$ and $b$-field gauge transformations 
parametrized by $\tilde{\xi}_i$. Conversely, keeping $\tilde{\partial}^i$ non-zero but setting 
$\partial_i=0$, the gauge transformations of DFT reduce in particular to general 
coordinate transformations in the dual coordinates, $\tilde{x}_i\rightarrow \tilde{x}_i-\tilde{\xi}_i(\tilde{x})$. 
Prior to solving the strong constraint by setting half of the derivatives to
zero, the full $\xi^M$ gauge symmetry is not manifest in terms of the conventional 
fields $g_{ij}$ and $b_{ij}$. In particular, the 
$\xi^{i}$ transformations act non-linearly, 
and originally the gauge invariance has 
only been verified through a lengthy computation \cite{Hohm:2010jy}.
Afterwards, a manifestly $O(10,10)$ invariant formulation has been found in \cite{Hohm:2010pp},
which linearizes the gauge transformations in terms of a generalized metric.  
This gives rise to a more geometrical formulation which,  
employing earlier work by Siegel \cite{Siegel:1993th}, has been further
developed in \cite{Hohm:2010xe,Hohm:2011si,Jeon:2011cn}. 
The geometrical formulations developed so far involve $O(10,10)$ covariant tensors that 
combine the metric $g$ and the $b$-field into a single object. Here, we will give   
a formulation in terms of the `component' fields $\tilde{g}$, $\beta$ and $\tilde{\phi}$
that makes half of the generalized diffeomorphisms, 
those parametrized by $\xi^{i}$, manifest.

In order to illustrate the problem that we would like to address, consider the DFT action in terms of the redefined fields which takes the 
schematic form 
 \bea\label{schDFT}
  S_{\rm DFT} = \int dx d\tilde{x} e^{-2d}\Big({\cal R}(\tilde{g},\partial)+{\cal R}(\tilde{g}^{-1},\tilde{\partial})+\cdots \Big)\;. 
 \eea 
Here, the first term denotes the conventional Ricci scalar ${\cal R}$ 
based on the metric $\tilde{g}_{ij}$. Similarly, the second term 
denotes the Ricci scalar with respect to winding derivatives $\tilde{\partial}^i$.
More precisely, the inverse metric $\tilde{g}^{ij}$ plays the role of the usual metric 
in order to work 
consistently with the upper indices of $\tilde{\partial}^i$.  
The first term is manifestly invariant under general coordinate transformations 
generated by $\xi^{i}$. Similarly, the second term is manifestly invariant under
general `winding' coordinate transformations generated by $\tilde{\xi}_i$.  
In DFT, however, the gauge parameters depend a priori on $x$ and $\tilde{x}$ and 
so the first term is not invariant under $\tilde{\xi}_i$ transformations and 
the second term is not invariant under $\xi^{i}$ transformations. 
In the following we will render the diffeomorphism symmetry parametrized 
by $\xi^{i}$ manifest by introducing a new connection that covariantizes  
the winding derivatives, $\tilde{\partial}^i\rightarrow \tilde{\nabla}^i$, 
which will naturally introduce $Q$ as the antisymmetric part 
of this connection.  

The action of the diffeomorphisms parametrized by $\xi^{i}$ 
can be read off from eq.~(2.36) of \cite{Hohm:2010jy}, 
 \bea\label{transform}
  \delta_{\xi}\tilde{g}_{ij} = {\cal L}_{\xi}\tilde{g}_{ij}\;, \quad \delta_{\xi}\beta^{ij}
  = \tilde{\partial}^{i}\xi^{j}-\tilde{\partial}^{j}\xi^{i}+{\cal L}_{\xi}\beta^{ij}\;, 
 \eea
where ${\cal L}_{\xi}$ denotes the Lie derivative, which acts in the 
usual way on tensors, 
 \bea
   {\cal L}_{\xi}\beta^{ij} =   \xi^{k}\partial_{k}\beta^{ij}-\beta^{ik}\partial_k \xi^{j}-\beta^{kj}\partial_{k}\xi^{i}\;, 
 \eea
and similarly on objects with an arbitrary index structure.    
We call a transformation covariant if it involves only 
the Lie derivative, and we denote the  
non-covariant part of a variation by $\Delta_{\xi}\equiv \delta_{\xi}-{\cal L}_{\xi}$, 
so that from (\ref{transform})
 \bea\label{Deltagb}
  \Delta_{\xi}\tilde{g}_{ij} = 0\;, \qquad \Delta_{\xi}\beta^{ij} =  \tilde{\partial}^{i}\xi^{j}-\tilde{\partial}^{j}\xi^{i}\;.
 \eea

Let us now develop a tensor calculus for the winding derivatives but with 
respect to the `momentum' diffeomorphisms generated by $\xi^{i}$. We will  define covariant 
derivatives and invariant curvatures. We start by considering a scalar like 
the dilaton $\tilde{\phi}$, whose tilde derivative transforms as 
 \be
 \begin{split}
  \delta_{\xi}(\tilde{\partial}^i\tilde{\phi})  =  \tilde{\partial}^i(\xi^{p}\partial_p\tilde{\phi}) 
   = \xi^{p}\partial_p(\tilde{\partial}^i\tilde{\phi})+\tilde{\partial}^i \xi^{p}\partial_p\tilde{\phi}\;. 
 \end{split}  
 \ee
In order to bring this into a form that is closer to the Lie derivative of 
a vector we add on the right-hand side 
 \be
  -\partial_p\xi^{i} \tilde{\partial}^p \tilde{\phi}-\tilde{\partial}^p\xi^i \partial_p\tilde{\phi} = 0\;, 
 \ee
which is zero due to the strong constraint (\ref{strng}), and obtain 
 \be
    \delta_{\xi}(\tilde{\partial}^i\tilde{\phi})  = {\cal L}_{\xi}(\tilde{\partial}^i\tilde{\phi}) + 
    (\tilde{\partial}^i \xi^{p}-\tilde{\partial}^p\xi^i) \partial_p\tilde{\phi}\;.
 \ee
Thus, $\tilde{\partial}^i\tilde{\phi}$ does not transform covariantly, but 
its non-covariant variation contains the same inhomogeneous term as 
the transformation (\ref{Deltagb}) of $\beta^{ij}$. Therefore, introducing  
the derivative operator 
 \be\label{Dtilde}
  \tilde{D}^{i} \equiv  \tilde{\partial}^i -\beta^{ij}\partial_{j}\;,   
 \ee
we see that $\tilde{D}^{i}\tilde{\phi}$ is a fully covariant 
derivative of a scalar.   
In the following, the derivative (\ref{Dtilde}) will play the role 
of a partial but anholonomic derivative. The $\tilde{D}^i$ 
are non-commuting, and their commutator reads 
   \be\label{commD}
   \big[ \tilde{D}^{i},\tilde{D}^{j}\big]  =  -R^{ijk}\partial_{k}-Q_{k}{}^{ij}\tilde{D}^{k}\;, 
  \ee
where 
 \be\label{Rflux}
  R^{ijk}  =  3\tilde{D}^{[i}\beta^{jk]} = 3\big(\tilde{\partial}^{[i}\beta^{jk]}+\beta^{p[i}\partial_{p}\beta^{jk]}\big) \;,
 \ee
and $Q$ is still given by (\ref{Q}). The proof of (\ref{commD}) requires the strong constraint (\ref{strng}).  
One may verify, using again (\ref{strng}), that $R^{ijk}$ transforms covariantly under (\ref{transform}). Thus, 
(\ref{Rflux}) represents the covariant field strength of $\beta^{ij}$ that we will
refer to as $R$ flux in the following and which coincides with the expression found in \cite{abmn11}.  

Next, we define derivatives that are covariant when acting on 
arbitrary tensors. For a vector $V$ we set 
 \be\label{covder}
  \tilde{\nabla}^{i}V^{j}  =  \tilde{D}^{i}V^{j}-\widecheck{\Gamma}_{k}{}^{ij}V^{k}\,, \quad
  \tilde{\nabla}^{i}V_{j}  =  \tilde{D}^{i}V_{j}+\widecheck{\Gamma}_{j}{}^{ik}V_{k}\,, 
 \ee   
and similarly for tensors with an arbitrary number of upper and lower indices.  
In order for (\ref{covder}) to transform covariantly, the connection $\widecheck{\Gamma}$
needs to transform as  
 \be\label{simpconn}
  \Delta_{\xi}\widecheck{\Gamma}_{k}{}^{ij}   =  -\tilde{D}^{i}\partial_k \xi^{j}\;.
 \ee
We note that the antisymmetric part $\widecheck{\Gamma}_{k}{}^{[ij]}$ 
does not transform as a tensor and therefore cannot be set to zero. 
In order to express the connection in terms of the physical fields 
we impose two covariant constraints. First, we require the metricity 
condition 
 \be
    \label{metricity}
    \tilde{\nabla}^i\tilde{g}^{jk}  =  0\;.
 \ee  
This determines the symmetric part $\widecheck{\Gamma}_{k}{}^{(ij)}$
in terms of the antisymmetric part
$\widecheck{\Gamma}_{k}{}^{[ij]}$ and $\tilde{D}^{i}\tilde{g}^{jk}$. 
The antisymmetric part in turn is naturally fixed by requiring that the commutator of 
covariant derivatives on a scalar is only given by the covariant term, involving the $R$ flux, 
 \bea
  \big[ \tilde{\nabla}^i,\tilde{\nabla}^j\big]\tilde{\phi} = -R^{ijk}\partial_k\tilde{\phi} \;.
 \eea
This implies with (\ref{commD}) 
 \bea
  \widecheck{\Gamma}_{k}{}^{[ij]} = -\frac{1}{2}Q_{k}{}^{ij}\;.
 \eea
Thus, $Q$ is given by the antisymmetric part of the
connection. The full connection is then
 \bea\label{fullconn}
  \widecheck{\Gamma}_{k}{}^{ij} =  \tilde{\Gamma}_{k}{}^{ij}+\tilde{g}_{kl}\tilde{g}^{p(i}Q_{p}{}^{j)l}-\frac{1}{2}Q_{k}{}^{ij}\;, 
 \eea
where 
 \be\label{tildeGamma}
  \tilde{\Gamma}_{k}{}^{ij} =  \frac{1}{2}\tilde{g}_{kl}\big(\tilde{D}^i\tilde{g}^{jl}+  \tilde{D}^j\tilde{g}^{il}-\tilde{D}^l\tilde{g}^{ij}\big)\;
 \ee        
are the conventional Christoffel symbols in the winding coordinates, 
but with $\tilde{\partial}^i$ replaced by $\tilde{D}^i$.    
The $R$ flux (\ref{Rflux}) satisfies a Bianchi identity that can be 
written in terms of these covariant derivatives as 
  \be
   \tilde{\nabla}^{[i}R^{jkl]}  = 0\;, 
  \ee
or, explicitly,  
 \be
  4\tilde{\partial}^{[i}R^{jkl]}+4\beta^{p[i}\partial_{p}R^{jkl]}+6Q_{p}{}^{[ij} R^{kl]p} =  0\;.  
 \ee
We finally note from (\ref{simpconn}) that the trace of the connection,  
 \be\label{newtorsion}
  {\cal T}^{i} \equiv  \widecheck{\Gamma}_{k}{}^{ki}\;,
 \ee  
transforms as a tensor,  
 \be
  \Delta_{\xi}{\cal T}^{i} = \Delta_{\xi}\widecheck{\Gamma}_{k}{}^{kj}  =  -\partial_{k}\tilde{\partial}^{k}\xi^{j}-\beta^{pk}\partial_p \partial_k\xi^{j} = 0\;, 
 \ee
by the strong constraint and the antisymmetry of $\beta$. 
This is analogous to the antisymmetric part of the conventional 
connection, i.e., the torsion tensor. 
Thus, we can think of ${\cal T}^i$ as a new torsion, and we 
stress that it is non-zero for (\ref{fullconn}).    

Having defined covariant derivatives  we next construct a 
Riemann tensor through the commutator of 
covariant derivatives, 
  \be\label{fullcomm}
   \big[ \tilde{\nabla}^{i},\tilde{\nabla}^{j}\big]V_{k}  =  -R^{ijp}\nabla_{p}V_{k}+ \widecheck{\cal R}^{ij}{}_{k}{}^{p}V_{p}\;, 
 \ee 
where    
 \be\label{RIEMANN}
 \begin{split}
  \widecheck{\cal R}^{ij}{}_{k}{}^{p}  =  &\,\tilde{D}^{i}\widecheck{\Gamma}_k{}^{jp}-\tilde{D}^{j}\widecheck{\Gamma}_{k}{}^{ip}
  +\widecheck{\Gamma}_{k}{}^{iq}\widecheck{\Gamma}_{q}{}^{jp}- \widecheck{\Gamma}_{k}{}^{jq}\widecheck{\Gamma}_{q}{}^{ip} \\
  &+Q_{q}{}^{ij}\widecheck{\Gamma}_k{}^{qp}-R^{ijq}\,\Gamma^{p}{}_{qk}\;.
 \end{split}
 \ee   
Here we have used the conventional covariant derivative for $\partial_{i}$, 
with Christoffel symbols $\Gamma^{k}{}_{ij}$ based on the metric 
$\tilde{g}_{ij}$.  As the $R$ flux is fully covariant, the two terms on the right-hand side of 
(\ref{fullcomm}) are separately covariant and therefore (\ref{RIEMANN}) 
defines a covariant curvature. From this we can define a Ricci tensor in the usual 
way, 
 \be
 \begin{split}
 \widecheck{\cal R}^{ij}  & \equiv \widecheck{\cal R}^{ki}{}_{k}{}^{j} \\
      &= \tilde{D}^{k}\widecheck{\Gamma}_{k}{}^{ij}-\tilde{D}^{i}\widecheck{\Gamma}_{k}{}^{kj}+
    \widecheck{\Gamma}_{k}{}^{ij}\widecheck{\Gamma}_{q}{}^{qk}-\widecheck{\Gamma}_{p}{}^{ki}\widecheck{\Gamma}_{k}{}^{pj} \\
    & = \tilde{D}^{k}\widecheck{\Gamma}_{k}{}^{ij}-\tilde{\nabla}^{i}{\cal T}^{j}
    -\widecheck{\Gamma}_{q}{}^{ki}\widecheck{\Gamma}_{k}{}^{qj}\;, 
 \end{split}   
 \ee
which in general will not be symmetric in $i,j$.  
Here, we used in the third equation that the trace of $\widecheck{\Gamma}$ yields 
the tensor (\ref{newtorsion}) with a well-defined covariant derivative. 
Thus, curiously, the Ricci tensor decomposes into two tensors that are separately 
covariant. Finally, we can define a Ricci scalar, 
  \bea
   \widecheck{\cal R} = \tilde{g}_{ij}\widecheck{\cal R}^{ij}\;.
  \eea
  
We are now ready to give as our main result the full DFT action for $\tilde{g}_{ij}$, $\beta^{ij}$ and $\tilde{\phi}$ 
in terms of the geometrical quantities defined above, 
 \be\label{conjaction}
 \begin{split}
  S_{\rm DFT}  =  \int   d&xd\tilde{x}\,\sqrt{-\tilde{g}}\,e^{-2\tilde{\phi}}\Big[{\cal R} + \widecheck{\cal R}-\frac{1}{12}R_{ijk} R^{ijk} \\
  &+4\Big( (\partial\tilde{\phi})^2
  + (\tilde{D}\tilde{\phi})^2 
  +\tilde{\nabla}^{i}{\cal T}_{i}-{\cal T}^{i}{\cal T}_{i}\Big) \Big] 
   \;. 
 \end{split}
 \ee     
This action is the precise version of the schematic form (\ref{schDFT}).  
It involves a kinetic term for the $R$ flux and two Einstein--Hilbert terms. 
The first is the conventional one 
for $\tilde{g}_{ij}$ based on the usual derivatives $\partial_{i}$.  The  
second one is based on the winding derivatives but involving the novel 
connection \eqref{fullconn} including the $Q$ flux. Moreover, the new torsion ${\cal T}^i$
is required in order to reproduce the full DFT. In (\ref{conjaction}) every term is 
manifestly invariant under the diffeomorphisms generated by $\xi^{i}$.                                
 
Upon setting $\tilde{D}^{i}=0$ and ${\cal T}^i=0$ we recover the situation analyzed in \cite{allp11},
and the $Q^2$ term in (\ref{Qaction}) is the only remnant left of the `dual' 
Einstein--Hilbert term. We may also set $\tilde{\partial}^i=0$ but keep $\tilde{D}^{i}=-\beta^{ij}\partial_{j}$
and ${\cal T}^i$, 
for which (\ref{conjaction}) reduces to the action obtained from 
the standard NS-NS action (\ref{original}) by performing the field 
redefinition (\ref{tildeE}) without the simplifying assumption. 
This ten-dimensional supergravity action contains the $R^2$ term, in which the $R$ flux 
is reduced to the second term in (\ref{Rflux}). It also 
contains $Q^2$ terms and various couplings of $\beta$ to dilaton and 
metric, but its geometric form and invariance is obscured in absence of
the winding derivatives. 

On a technical level the action (\ref{conjaction}) provides 
an alternative formulation of DFT that makes half of the gauge 
symmetries, those parametrized by $\xi^{i}$, manifest. 
The remaining gauge symmetries spanned by $\tilde{\xi}_i$
are hidden in the formulation (\ref{conjaction}), but we may 
return to the original fields $g_{ij}$ and $b_{ij}$  
and employ the `T-dual' of the geometrical structures discussed 
here. The $R$ flux is then replaced by a covariantized 
$H$ flux,  
 \bea
  H_{ijk} = 3\big(\partial_{[i}b_{jk]}+b_{p[i}\tilde{\partial}^{p}b_{jk]}\big)\;, 
 \eea
and similarly all other objects result from those introduced
here by replacing the fields by the original fields,  
sending $\partial_{i}\leftrightarrow \tilde{\partial}^i$
and, generally, consistently interchanging upper with lower indices. 
The resulting action will then be manifestly invariant under 
$\tilde{\xi}_i$ gauge transformations. Thus, for each half 
of the gauge transformations there is a field basis in which this 
symmetry can be made manifest. In total this yields an alternative 
proof of the full gauge  
invariance of DFT.

One can perform a Kaluza-Klein reduction of supergravity written in 
the new variables in order to make contact with the 
non-geometric fluxes (\ref{chain}) in four dimensions. 
In practice one would then redefine only the field components along the 
internal directions. 
This indeed leads to scalar potential terms containing $Q$ and $R$
fluxes of the required form, but without $H$ flux \cite{ahllp12}. 
The gauged supergravities thus obtained are, however, not the most general 
ones; they contain those related via T-duality to gaugings 
without non-geometric fluxes.  
One purpose of the field redefinitions in ten dimensions is to obtain
a background solution that is globally 
well-defined and leads to a proper 
Kaluza-Klein reduction, thereby providing a higher-dimensional origin of
gaugings that are T-dual to geometric ones. 
Our results are consistent with the findings of 
 \cite{abmn11, geiss} showing that the most general ${\cal N}=4$
gauged supergravities in four dimensions result  
from DFT only if the strong constraint is relaxed. 
Recently, mild relaxations of this constraint have been found to be 
consistent \cite{Hohm:2011cp,Grana:2012rr}. 
In this letter we worked with the strongly constrained DFT, 
but we hope that the geometrical structures 
found here will provide a guide for 
non-geometric compactifications more generally.

Finally, the action constructed here 
may also be closely related to the effective
action of non-commutative and non-associative gravity, which describes  
closed strings on backgrounds with non-geometric fluxes \cite{Blumenhagen:2010hj}.

\begin{acknowledgments} \vspace{-0.25cm}
We would like to thank Barton Zwiebach for comments. 
This work is supported by the Alexander-von-Humboldt foundation, the
DFG Transregional Collaborative Research Centre TRR 33
and the DFG cluster of excellence `Origin and Structure of the Universe'.
DL thanks the Simons Center for Geometry and Physics, and ML thanks the Isaac 
Newton Institute in Cambridge, as part of the programme on the Mathematics and Applications of Branes in String and M-theory,  for hospitality.
\end{acknowledgments}

\end{document}